\documentclass[submission,copyright,creativecommons]{eptcs}
\providecommand{\event}{GASCom 2024} 

\pdfoutput=1
\usepackage{iftex}

\ifpdf
  \usepackage{underscore}         
  \usepackage[T1]{fontenc}        
\else
  \usepackage{breakurl}           
\fi

\usepackage{graphicx}
\usepackage{amssymb}
\usepackage{amsmath}
\usepackage{amsthm}
\usepackage{amsfonts} 
\usepackage{amscd}
\usepackage{booktabs}
\usepackage{multirow}
\usepackage{mathtools}
\usepackage{subcaption}

\usepackage{tikz}
\usetikzlibrary{automata}
\usepackage{tikz-cd}
\usepackage{enumitem}
\usetikzlibrary{calc, positioning, fit, shapes.misc}
\usepackage{cleveref}

\definecolor{Madder}{HTML}{A31621}
\definecolor{Sage}{HTML}{C5C9A4}
\definecolor{TeaRose}{HTML}{ECBEB4}
\definecolor{RoseQuartz}{HTML}{9C92A3}

\newcommand{\calH}{\mathcal{H}}

\theoremstyle{plain}

\newtheorem{theorem}{Theorem}
\newtheorem{corollary}{Corollary}
\newtheorem{problem}{Problem}
\newtheorem{prop}{Proposition}

\theoremstyle{definition}
\newtheorem{definition}{Definition}

\theoremstyle{remark}

\newtheorem{example}{Example}

\newcommand{\ssq}{
	\vcenter{\hbox{\scalebox{0.65}{$\;\mathbin{ \square }\;$}}}
}

\title{Combinatorics on Social Configurations}
\author{Dylan Laplace Mermoud
\institute{UMA, ENSTA Paris, Institut Polytechnique de Paris, \\ Paris, France.}
\email{dylan.laplace@ensta-paris.fr}
\and
Pierre Popoli 
\institute{Department of Mathematics, ULiège \\ Liège, Belgium.}
\email{Pierre.Popoli@uliege.be}
}
\def\titlerunning{Combinatorics on Social Configurations}
\def\authorrunning{D. Laplace Mermoud \& P. Popoli}

\begin{document}

\sloppy

\maketitle

In cooperative game theory, the social configurations of players are modeled by balanced collections~\cite{bondareva1963some, shapley1967balanced}. A balanced collection is a set system defined on the set \(N\) of players in the game, together with a system of weights such that each player belongs to coalitions whose weights sum to \(1\). The Bondareva--Shapley theorem, perhaps the most fundamental theorem in cooperative game theory, characterizes the existence of solutions to the game that benefit everyone using balanced collections. Roughly speaking, if the trivial set system \( \{ N \} \) is one of the most efficient balanced collections for the game, then the set of solutions from which each coalition benefits, the so-called \emph{core}, is non-empty.

\smallskip 

In the following, we discuss some interactions between combinatorics and cooperative game theory that are still relatively unexplored. First, we study the similarities between balanced collections on the one hand and regular or uniform hypergraphs on the other. Second, we present some results leading to the construction of the combinatorial species of structures of uniform hypergraphs, from which we aim to construct the species of regular hypergraphs by duality. Finally, we investigate the possibility of expressing some ``minimality'' properties of regular or uniform hypergraphs in the language of combinatorial species, hoping to obtain new properties of minimal balanced collections.

\section{Cooperative Game Theory}

Cooperative game theory aims to study the emergence of cooperative behavior between rational players whose actions affect each other's well-being. It was introduced in the seminal book \emph{Theory of Games and Economic Behavior} by von Neumann and Morgenstern~\cite{von1944theory}, written during the Second World War, motivated by von Neumann's desire to study the stability of social organizations.

\begin{definition}[von Neumann and Morgenstern~\cite{von1944theory}]
    A \emph{cooperative game with transferable utility}, hereafter called \emph{game}, is an ordered pair \( (N, v) \) where
    \begin{itemize}
        \setlength\itemsep{-0.1em}
        \item \(N\) is a non-empty finite set of \emph{players}, called the \emph{grand coalition},
        \item \(v\) is a set function \( v: 2^N \to \mathbb{R} \) such that \( v(\emptyset) = 0 \). 
    \end{itemize}
\end{definition}

The non-empty subsets of \( N \) are called \emph{coalitions}, and their set is denoted by \( \mathcal{N} \). For each coalition \( S \in \mathcal{N} \), the number \( v(S) \), called the \emph{worth} of \(S\), can be interpreted as the amount of \emph{utility} or \emph{satisfaction} that the players forming \(S\) can obtain through full cooperation. When a coalition is formed, a non-trivial task is to allocate among its players the utility acquired by the coalition among its players. To prevent the coalition from splitting, the allocation of each of its subcoalitions must at least pay off its value, otherwise the coalitions would defect to obtain more utility. A necessary condition for the formation of the grand coalition is therefore that the following set
\[
C(v) = \left\{x \in \mathbb{R}^N \Big| \; \sum_{i \in N} x_i = v(N), \text{ and } \sum_{i \in S} x_i \geq v(S), \forall S \in \mathcal{N} \right\}
\]
is not empty. The set \( C(v) \) is called the \emph{core} of the game and is one of the essential objects studied in cooperative game theory. Each vector \( x \in \mathbb{R}^N \) represents a payment to the players when player \(i \in N\) receives a payment from \(x_i\). The payment of a coalition is the sum of the payments of its players. Thus, the vectors in the core are exactly the payments that allocate the utility acquired by the large coalition in such a way that each coalition is satisfied with its payment.

\smallskip 

A closely related object is the balanced collection. Formally, a balanced collection \( \mathcal{B} \) is a set of coalitions such that there exists a map \( \lambda: \mathcal{B} \to \mathbb{R}_{> 0} \) satisfying \( \sum_{S \in \mathcal{B}, S \ni i} \lambda(S) = 1 \) for each player \( i \in N \). For example, the set partitions of \( N \) are balanced collections with unit weights. We measure the efficiency of a balanced collection \( \mathcal{B} \) by taking the weighted sum of the worths of the coalitions in \( \mathcal{B} \), that is \(\sum_{S \in \mathcal{B}} \lambda(S) v(S)\).

\begin{theorem}[Bondareva~\cite{bondareva1963some}, Shapley~\cite{shapley1967balanced}]
    The core of a game is nonempty if and only if \( \{N\} \) belongs to the set of maximally efficient balanced collections.
\end{theorem}

The Bondareva-Shapley theorem provides a useful characterization of the core nonemptiness, from which the first author, Grabisch and Sudh{\"o}lter~\cite{laplace2023minimal} developed an algorithm. This algorithm is based on an improved characterization of core nonemptiness, often called the sharp Bondareva-Shapley theorem, which differs from the previously mentioned theorem only in that the balanced collections are replaced by the \emph{minimal balanced collections}. The minimal balanced collections are the balanced collections for which no proper subcoalitions are balanced. Moreover, the set of minimal balanced collections is the minimal, with respect to inclusion, set of balanced collections for which the Bondareva-Shapley theorem holds. In the same paper, the first author, Grabisch and Sudh{\"o}lter~\cite{laplace2023minimal} have generated the minimal balanced collections up to \(7\) players. The sequence of the numbers of the minimal balanced collections is stored as \href{https://oeis.org/A355042}{\underline{A355042}} in the \emph{Online Encyclopedia of Integer Sequences}~\cite{oeis}. The method used in the aforementioned paper is inefficient when the number of players is greater than \(7\), and this work aims to find another way to generate it.

\begin{table}[ht]
\begin{center}
\begin{tabular}{lcccccccc}
\toprule
$n$ \hspace{0.5cm} & \hspace{3pt} 2 \hspace{3pt} & \hspace{3pt} 3 \hspace{3pt} & \hspace{3pt} 4 \hspace{3pt} & \hspace{3pt} 5 \hspace{3pt} & \hspace{18pt} 6 \hspace{18pt} & \hspace{3pt} 7 \hspace{3pt} \\ 
\midrule
$k$ & 2 & 6 & 42 & 1,292 & 200,214 & 132,422,036 \\
\bottomrule
\end{tabular}
\caption{Number $k$ of minimal balanced collections according to the number $n$ of players.}
\label{table: perf}
\end{center}
\end{table}

\vspace*{-30pt}

\section{Hypergraphs}

The cornerstone of this work is the striking similarity between the balanced collections and the regular hypergraphs. An \emph{(undirected) hypergraph} \( \mathcal{H} \) is a pair \( \mathcal{H} = (N, E) \), where \(N\) is a set of \emph{nodes} and \(E\) is a spanning collection of non-empty subsets of \(N\), called \emph{hyperedges} or simply \emph{edges}. 

\smallskip 

A hypergraph \( \mathcal{H} \) is called \emph{k-regular} if for each node \( x, \in N\) the \emph{degree} of $x$ is $k$, i.e. \( \delta(x) \coloneqq \lvert \{e \in E \mid e \ni x \} \rvert=k\). The underlying set of the multiset of edges of a regular hypergraph is a balanced collection. Indeed, the weight of a given coalition is the multiplicity of the edge in the collection \(E\) divided by the regularity of the hypergraph. If each edge has cardinality $d$, the hypergraph is said to be \emph{d-uniform}. Therefore, the dual of a \(d\)-regular hypergraph is \(d\)-uniform and vice versa. 

\smallskip 

One of the main interests of uniform hypergraphs lies in the fact that writing a program that generates uniform hypergraphs of a certain size, i.e. with a certain number of edges, is extremely simple. It is sufficient to take arbitrary sets of equal cardinality and relabel their elements so that they fit into the notation \(N = \{1, \ldots, n\}\). If an edge needs to be added to the hypergraph, any set of nodes with the appropriate cardinality can be used. However, if we want to add a node in a regular hypergraph, it is not easy to add it while maintaining regularity. Note that adding an edge to a uniform hypergraph is the same operation as adding a node to its dual regular hypergraph.

\smallskip 

We believe that this approach is a possible route to a more efficient method for generating minimal balanced collections. Let \( \mathcal{H} = (N, E) \) be a hypergraph, let \( A \subseteq N \) and \( X \subseteq E \). The hypergraph denoted by \( \mathcal{H}_A \) and defined by \(\mathcal{H}_A = \left(A, \left\{ S \cap A \mid S \in E \right\} \right)\) is the \emph{subhypergraph} of \( \mathcal{H} \) induced by \(A\). The hypergraph \(\mathcal{H}^X = (N, X) \) is the \emph{partial hypergraph} of \( \mathcal{H} \) induced by \(X\). Note that the subhypergraph of a hypergraph corresponds to a partial hypergraph of its dual.

\smallskip 

Similarly to minimal balanced collections, we say that a hypergraph is \emph{minimally uniform} if it is uniform and no proper subhypergraph is uniform, and we say that a hypergraph is \emph{minimally regular} if it is regular and no proper partial hypergraph is regular.

\begin{prop}\label{prop: dual}
 The dual of a minimally uniform hypergraph is minimally regular and vice versa.
\end{prop}

Note that our definition of a subhypergraph does not eliminate the edges that become empty when taking the intersection with the subset of nodes. This definition is not common in the literature, but it is a natural one in our context, and the proposition above illustrates this fact.

\section{Species of structures}

Our goal now is to generate the objects we mentioned. To do this, we use the theory of \emph{species of structures} and the corresponding operations on formal power series developed by Joyal~\cite{joyal1981theorie}.

\smallskip

A species of structures is a rule \( \text{F} \) that assigns to each finite set \(U\) a finite set \( \text{F}[U] \) that is ``independent of the nature'' of the elements of \( U \). The members of \( \text{F}[U] \), called \emph{\( \mathrm{F} \)-structures}, are interpreted as combinatorial structures on the set \(U\) given by the rule \(\text{F}\). The fact that the rule is independent of the nature of the elements of \(U\) is expressed by the invariance under relabeling. More precisely, to any bijection \(\sigma: U \to V\) the rule \(\text{F}\) associates a bijection \(\text{F}[\sigma]: \text{F}[U] \to \text{F}[V]\) that transforms each \(\mathrm{F}\)-structures on \(U\) into an (isomorphic) \(\text{F}\) structure on \(V\).

\smallskip 

Each species is associated with a formal power series, which refers to the enumeration of \(\mathrm{F}\) structures and is denoted by \(\mathrm{F}(x)\). There are a myriad of operations on species of structures such as addition, multiplication, functorial and partitional composite, see \cite{bergeron1998combinatorial} for more details and further operations. The main interest of these operations is to provide a new description of a species of structures and to extract formulas over the generating series.

\begin{example}

Let \(\wp\) denote the species of subsets associating to each finite set \(U\) the set of subsets of \(U\), and \( \wp^{ [2]} \) the species of the \(2\)-subsets, or unordered pairs, defined similarly. Their generating series are, respectively, \(\wp^{ [2]}(x) = \sum_{n \geq 0} \binom{n}{2} \frac{x^n}{n!} \) and \(\wp(x) = \sum_{n \geq 0} 2^n \frac{x^n}{n!} = e^{2x}\). Thanks to these two species and the composition of species, we have the following combinatorial identity \begin{align} \label{eq:graph}
\textsc{Gr}=\wp \ssq \wp^{ [2]} \end{align} where \(\textsc{Gr}\) is the species of simple graphs. From this formula, we obtain the generating series of simple graph, namely \(\textsc{Gr}(x)=\sum_{n \geq 0} 2^{\binom{n}{2}}\frac{x^n}{n!}\). An illustration of this identity is pictured in Figure~\ref{fig:graph}.

\begin{figure}[!h] 
\begin{center}
\begin{subfigure}{0.45\textwidth}
\begin{center}
\begin{tikzpicture}[thick, main/.style = {draw, circle, inner sep=1pt, fill}] 
\node[main] (4) at (1, 0) {};
\node[main] (2) at (0.3, 0.2) {};
\node[main] (3) at (0.8, 1) {};
\node[main] (1) at (0.1, 0.8) {};
\node[main] (5) at (-0.5, 0.5) {};
\node[main] (6) at (-0.3, -0.2) {};
\node[main] (7) at (-0.25, -1) {};
\node[main] (8) at (0, -0.5) {};
\node[main] (9) at (-0.5, -0.75) {};

\draw[teal] (6) to (4);
\end{tikzpicture} 
\subcaption{A typical element of $\wp^{[2]}$.}
\end{center}
\end{subfigure}
\hspace{0.1cm}
\begin{subfigure}{0.45\textwidth}
\begin{center}
\begin{tikzpicture}[thick, main/.style = {draw, circle, inner sep=1pt, fill}] 
\node[main] (4) at (1, 0) {};
\node[main] (2) at (0.3, 0.2) {};
\node[main] (3) at (0.8, 1) {};
\node[main] (1) at (0.1, 0.8) {};

\node[main] (5) at (-0.5, 0.5) {};
\node[main] (6) at (-0.3, -0.2) {};

\node[main] (7) at (-0.25, -1) {};
\node[main] (8) at (0, -0.5) {};
\node[main] (9) at (-0.5, -0.75) {};

\draw[teal] (6) to (4);
\draw[teal] (6) to (2);
\draw[teal] (4) to (1);
\draw[teal] (6) to (8);
\draw[teal] (7) to (9);
\draw[teal] (1) to (3);
\draw[teal] (2) to (8);
\end{tikzpicture} 
\subcaption{A typical element of $\wp \ssq \wp^{[2]}$.}
\end{center}
\end{subfigure}
\caption{Construction of the species \textsc{Gr} of simple graphs.}
\label{fig:graph}
\end{center}
\end{figure}
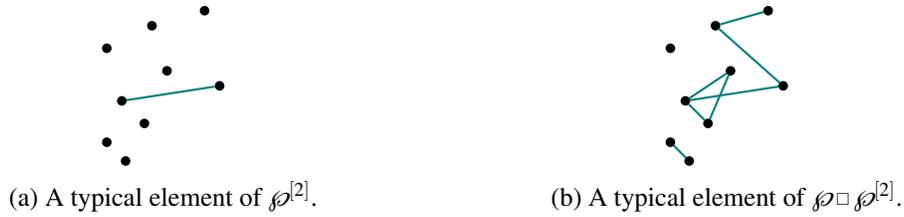
\end{example}

Let us denote $E$ the species of sets and \(\varsigma^{[p]}\) the species of \(k\)-subsets. Similarly to the combinatorial identity (\ref{eq:graph}), we have proved the following formula.

\begin{theorem}  \label{thm:unif}
The species of $k$-uniform hypergraphs of size $p$, which we denote by ${\normalfont \textsc{Uni}}_{k,p}$, satisfies the following combinatorial equation:
\[
{\normalfont \textsc{E}} \cdot {\normalfont \textsc{Uni}}_{k,p} = \varsigma^{[p]} \ssq \wp^{[k]}. 
\]
\end{theorem}

Let us denote \(n^{\overline{p}}=n(n+1)\cdots (n+p)\). Using the formalism of virtual species, see~\cite{bergeron1998combinatorial} again for more details, we have the following corollary. 

\begin{corollary}
The generating series of the species ${\normalfont \textsc{Uni}}_{k,p}$ is
\[
{\normalfont \textsc{Uni}}_{k,p}(x) = \sum_{n \geq 0} \left( \sum_{i = 0}^n (-1)^{n-i} \binom{n}{i} \frac{\binom{i}{k}^{\overline{p}}}{p!} \right) \frac{x^n}{n!}. 
\]
\end{corollary}

\begin{example}
Let us count the $2$-uniform hypergraphs of size $3$, with no more than three nodes. Since the hypergraphs are $2$-uniform, $n$ only goes from $2$ to $3$. Note that the number of hypergraphs is not counted up to an isomorphism. The number we are looking for is therefore
\[ \begin{aligned} 
\sum_{n = 2}^3 \left( \sum_{i = 2}^n (-1)^{n-i} \binom{n}{i} \frac{\binom{i}{2}^{\overline{p}}}{p!} \right) & = (-1)^0 \binom{2}{2} \frac{\binom{2}{2}^{\overline{3}}}{3!} + (-1)^1 \binom{3}{2} \frac{\binom{2}{2}^{\overline{3}}}{3!} + (-1)^0 \binom{3}{3} \frac{\binom{3}{2}^{\overline{3}}}{3!} \\
& = 1 - 3 + 10 = 8.
\end{aligned} \]

We represent them in the following. Notice that among these $8$ uniform hypergraphs, only one is minimal, that is the triangle. 

\begin{figure}[!h]
\centering
\begin{tikzpicture}
\tikzstyle{every node}=[shape=circle,fill=none,draw=black,minimum size=10pt,inner sep=2pt]
\tikzstyle{every path}=[color =black, line width = 0.5 pt]

\node(a0) at (0,0) {};
\node(b0) at (2,0) {};

\draw [bend left]  (a0) to (b0);
\draw [bend right] (a0) to (b0);
\draw [-] (a0) to (b0);


\node(a1) at (0,-2) {};
\node(b1) at (2,-2) {};
\node(c1) at (1,-1) {};

\draw [-] (a1) to (b1);
\draw [-] (b1) to (c1);
\draw [-] (c1) to (a1);


\node(a2) at (3,0) {};
\node(b2) at (5,0) {};
\node(c2) at (4,1) {};

\draw [bend right] (a2) to (b2);
\draw [bend left] (a2) to (b2);
\draw [-] (a2) to (c2);

\node(a3) at (6,0) {};
\node(b3) at (8,0) {};
\node(c3) at (7,1) {};

\draw [bend right] (b3) to (c3);
\draw [bend left] (b3) to (c3);
\draw [-] (a3) to (c3);

\node(a6) at (9,0) {};
\node(b6) at (11,0) {};
\node(c6) at (10,1) {};

\draw [bend right] (a6) to (c6);
\draw [bend left] (a6) to (c6);
\draw [-] (a6) to (b6);


\node(a4) at (3,-2) {};
\node(b4) at (5,-2) {};
\node(c4) at (4,-1) {};

\draw [bend right] (a4) to (b4);
\draw [bend left] (a4) to (b4);
\draw [-] (b4) to (c4);

\node(a5) at (6,-2) {};
\node(b5) at (8,-2) {};
\node(c5) at (7,-1) {};

\draw [bend right] (b5) to (c5);
\draw [bend left] (b5) to (c5);
\draw [-] (b5) to (a5);

\node(a7) at (9,-2) {};
\node(b7) at (11,-2) {};
\node(c7) at (10,-1) {};

\draw [bend right] (a7) to (c7);
\draw [bend left] (a7) to (c7);
\draw [-] (b7) to (c7);
;
\end{tikzpicture}
\caption{All \(2\)-uniform hypergraphs of size \(3\) with no more than \(3\) nodes.}
\end{figure}
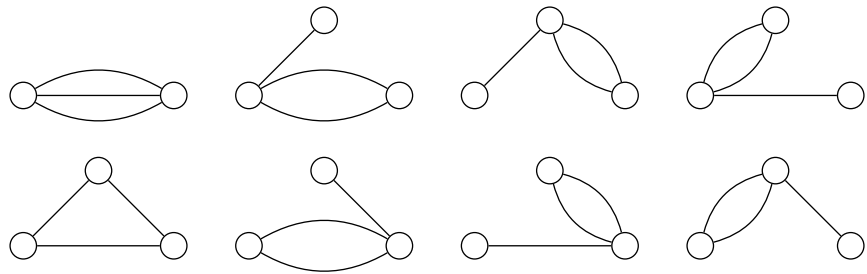

\end{example}

One can define the species of minimal balanced collections which are the underlying sets of the multisets of edges of minimally regular hypergraphs, which we construct from the minimally uniform hypergraphs, thanks to Proposition~\ref{prop: dual}. For now, we simply have constructed the species of uniform hypergraphs. 

\begin{problem}
 \label{prob:minimality} Express minimality in terms of species of structures.
\end{problem}

\section{Decompositions in minimally uniform hypergraphs}


Our first approach to study Problem~\ref{prob:minimality} was based on the idea that minimally uniform hypergraphs (resp. minimally regular hypergraphs) are the building blocks of uniform hypergraphs (resp. regular hypergraphs). In this work, we also assume that the number of edges remains the same since the goal is that they represent the number of players, and this should be fixed. The following proposition states that a uniform hypergraph can be partitioned into smaller minimally uniform hypergraph.

\begin{prop}
 Let $\calH = (N, E)$ be a uniform hypergraph of size $p$. Then there exists a partition \( \pi \) of \( N \) such that for each element \( B \in \pi\) the subhypergraph \( \calH_B \) is minimally uniform of size $p$.
\end{prop}

One can expect that the aforementioned partition is unique, up to a permutation, and therefore leads to a combinatorial identity via structures of species. However, the next example shows that such a partition is not unique. 

\begin{example}

Let us consider \( \mathcal{H} \) the \(4\)-uniform hypergraph of order \(7\) and size \(4\) defined by 
\[
\mathcal{H} = \left( \{v_1,\ldots,v_7\}, \big\{ \{v_1,v_2,v_3,v_4\}, \{v_1,v_5,v_6,v_7\}, \{v_3,v_4,v_5,v_6\}, \{v_3,v_4,v_6,v_7\} \big\} \right). 
\] The hypergraph \( \mathcal{H} \) can be partitioned in the two following ways

\begin{figure}[h]

\centering
\begin{tikzpicture}[node distance={5mm}, label distance=2mm, thick, main/.style = {draw, circle, inner sep=1pt, fill}] 

\tikzstyle{vertex} = [fill, shape=circle, node distance=55pt]
\tikzstyle{edge} = [fill, opacity=.5, fill opacity=.5, line cap=round, line join=round, line width=25pt]
\tikzstyle{elabel} =  [fill, shape=circle, node distance=50pt]

\draw[RoseQuartz, ultra thick, fill=RoseQuartz!80,opacity=0.6,rounded corners](-0.5,0.25) -- (6.25,0.25) -- (6.25,-0.25) -- (-0.5,-0.25) -- cycle; 
\draw[Sage, ultra thick, fill=Sage!80,opacity=0.6,rounded corners] (0,0.25) -- (2,-0.75) -- (6.35,-0.75) -- (6.35,-1.25) -- (1.75,-1.25) -- (-0.35,0) -- cycle; 
\draw[Madder, ultra thick, fill=Madder!80,opacity=0.6,rounded corners] (1.15,-1.25) -- (4,0.25) -- (6.5,0.25) -- (4.15,-1.25) -- cycle; 
\draw[TeaRose, ultra thick, fill=TeaRose!80, opacity=0.6,rounded corners] (3.75,0.25) -- (6.25,0.25) -- (6.25,-1.25) -- (3.75,-1.25) -- cycle; 

\node[main, label=above:{$v_1$}] (1) at (0,0) {};
\node[main, label=above:{$v_2$}] (2) at (2, 0) {};
\node[main, label=above:{$v_3$}] (3) at (4, 0) {};
\node[main, label=above:{$v_4$}] (4) at (6, 0) {};
\node[main, label=below:{$v_5$}] (5) at (2, -1) {};
\node[main, label=below:{$v_6$}] (6) at (4, -1) {};
\node[main, label=below:{$v_7$}] (7) at (6, -1) {}; 

\draw [-{Stealth[length=5mm]}] (2,-2) -- (0,-2.75);

\node[main, label=above:{$v_1$}] (11) at (-4,-4) {}; 
\node[main, label=above:{$v_3$}] (31) at (-2, -4) {};
\node[main, label=below:{$v_6$}] (61) at (-3, -5) {};

\draw [-,color=RoseQuartz] (11) to (31);
\draw [-,color=Sage] (11) to (61);
\draw [-,color=Madder,bend left] (61) to (31);
\draw [-,color=TeaRose,bend right] (61) to (31);

\node[main, label=above:{$v_2$}] (21) at (-0.5, -4) {};
\node[main, label=above:{$v_4$}] (41) at (1.5, -4) {};
\node[main, label=below:{$v_5$}] (51) at (-0.5, -5) {};
\node[main, label=below:{$v_7$}] (71) at (1.5, -5) {}; 

\draw [-,color=RoseQuartz] (21) to (41);
\draw [-,color=Sage] (51) to (71);
\draw [-,color=Madder] (41) to (51);
\draw [-,color=TeaRose] (41) to (71);

\draw [loosely dashed] (2.75,-3.25) -- (2.75,-5.75);

\draw [-{Stealth[length=5mm]}] (3.5,-2) -- (5.5,-2.75);

\node[main, label=right:{$v_2$}] (22) at (4,-4) {}; 
\node[main, label=below:{$v_6$}] (62) at (4, -5) {};

\tikzset{every loop/.style={min distance=10mm}}

\draw [loop,color=RoseQuartz] (22) to (22);
\draw [loop above,color=Sage] (62) to (62);
\draw [loop right,color=Madder] (62) to (62);
\draw [loop left,color=TeaRose] (62) to (62);

\draw[RoseQuartz, ultra thick, fill=RoseQuartz!80,opacity=0.6,rounded corners] 
(5.5,-3.75) -- (10.65,-3.75) -- (10.65,-4.25) -- (5.5,-4.25) -- cycle; 

\draw[Sage, ultra thick, fill=Sage!80,opacity=0.6,rounded corners] (5.5,-3.75) -- (6.5,-5.25) -- (9.5,-5.25) -- (9.5,-4.75) -- (6.85,-4.75) -- (6.25,-3.75) -- cycle; 

\draw[Madder, ultra thick, fill=Madder!80,opacity=0.6,rounded corners] (7.5,-3.75) -- (10.5,-3.75) -- (10.5,-4.25) -- (7.85,-4.25) -- (7.25,-5.25) -- (6.5,-5.25) -- cycle; 

\draw[TeaRose, ultra thick, fill=TeaRose!80, opacity=0.6,rounded corners] (7.25,-3.7) -- (10.75,-3.7) -- (9.25,-5.25) -- (8.75,-5.25) -- cycle; 

\node[main, label=above:{$v_1$}] (12) at (6, -4) {};
\node[main, label=above:{$v_3$}] (32) at (8, -4) {}; 
\node[main, label=above:{$v_4$}] (42) at (10, -4) {};
\node[main, label=below:{$v_5$}] (52) at (7, -5) {};
\node[main, label=below:{$v_7$}] (72) at (9, -5) {}; 

\end{tikzpicture}
\caption{Example of a non-unique partition.}
\end{figure}
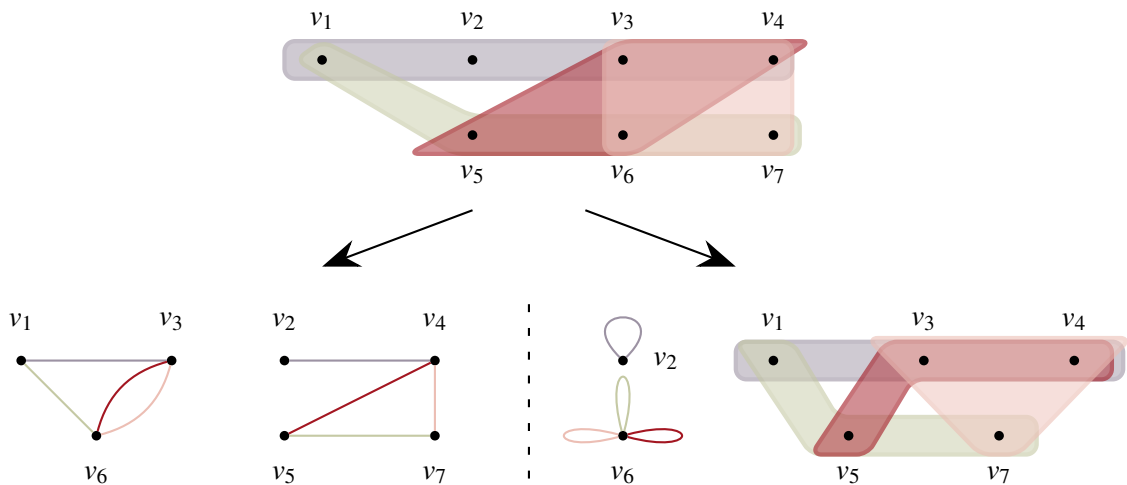

We can easily verify that \begin{itemize}
 \item The two hypergraphs on the left side \( \mathcal{H}_1 = \big( \{v_1,v_3,v_6\}, \big\{ \{v_1,v_3\}, \{v_1,v_6\}, \{v_3,v_6\}, \{v_3,v_6\} \big\} \big) \) and \( \mathcal{H}_2 = \big( \{v_2,v_4,v_5,v_7\}, \big\{ \{v_2,v_4\}, \{v_4,v_5\}, \{v_4,v_7\}, \{v_5,v_7\} \big\} \big) \) are minimally $2$-uniform hypergraphs that will merge to \( \mathcal{H} \).
 \item The hypergraphs on the right side \( \mathcal{H}'_1 = \big( \{v_2,v_6\}, \big\{ \{v_2\}, \{v_6\}, \{v_6\}, \{v_6\} \big\} \big) \) and \( \mathcal{H}'_2 = \big( \{v_1,v_3,v_4,v_5,v_7\}, \big\{ \{v_1,v_3,v_4\}, \{v_1,v_5,v_7\}, \{v_3,v_4,v_5\}, \{v_3,v_4,v_7\} \big\} \big) \) are respectively minimally $1$-uniform and minimally $3$-uniform hypergraphs that will merge to \( \mathcal{H} \).
\end{itemize}
\end{example}

Therefore, such a decomposition in the state cannot lead to a combinatorial identity such as \Cref{thm:unif}, which contains the species of minimally uniform hypergraph.

\bibliographystyle{eptcs}
\bibliography{generic}
\end{document}